\documentclass[aps,prl,showpacs,showkeys,groupedaddress]{revtex4}
\usepackage{amsmath,amsfonts,amssymb,amscd,amsthm,amsbsy,upref}
\numberwithin{equation}{section}

\def\F{{\mathcal F}}
\def\H{{\mathcal H}}
\def\O{{\mathcal O}}
\def\cee{{\mathbb C}}
\def\real{{\mathbb R}}
\def\zed{{\mathbb Z}}
\def\cl{\operatorname{cl}}
\def\Re{\operatorname{Re}}
\def\defeq{\ \buildrel {\text{def}}\over =\ }
\def\dag{\dagger}
\begin{document}
\title{Long time Evolution of Quantum Averages\\
Near Stationary Points}
\author{Gennady Berman} 
\affiliation{Los Alamos National Laboratory, MS B213\\
Los Alamos, NM 87545, U.S.A.}
\email[]{gpb@lanl.gov}
\author{Misha Vishik}
\affiliation{Department of Mathematics\\ The University of Texas at Austin\\
Austin, TX 78712-1082, U.S.A.}
\email[]{vishik@math.utexas.edu}
\date{\today}
\begin{abstract}
We construct explicit expressions for quantum averages in coherent states 
for a Hamiltonian of degree~4 with a hyperbolic stagnation point. 
These expressions are valid for all times and ``collapse'' (i.e., become 
infinite) along a discrete sequence of times. 
We compute quantum corrections compared to classical expressions. 
These corrections become significant over a time period of order 
$C\log \frac1{\hbar}$.
\end{abstract}


\maketitle

\baselineskip=18pt		

The central result of this paper is the exactly solvable evolution of 
quantum averages in coherent state for a Hamiltonian of degree~4 containing 
a hyperbolic point. 
We tend to think about this Hamiltonian 
$$H(a^\dag ,a) = i\omega (a^{\dag 2} - a^2) + \mu (a^{\dag 2} - a^2)^2$$
as a model one giving a useful insight into the global in time evolution 
of quantum averages for a more general Taylor expansion around a hyperbolic 
point, where the explicit expression is hard to obtain. 
As far as we know this is the first explicit computation in the presence 
of a hyperbolic point (apart from the quadratic case which is classical). 
General properties of spreading of a quantum wave packet were considered 
in \cite{CR}.
In \cite{Bl} the quantum energy levels were computed, e.g., near a 
nondegenerate local maximum of a double-well potential (this corresponds 
to a hyperbolic point of the Hamiltonian). 

There are two qualitative conclusions from the solutions we obtained below. 
First, the quantum corrections near a hyperbolic point become of order~1 
over a logarithmic time $c\log \frac1{\hbar}$. 
This time has appeared in \cite{BZ}. 
(See also \cite{BeB}, \cite{Ch}.) 
Second, it turns out, for such a Hamiltonian quantum averages in coherent 
states do become infinite along a certain discrete sequence of times.
For example, given the observable $\hat x^2$, these singularities occur 
at $t= \frac{\pi}{32\mu\hbar} + \ell \frac{\ell}{16\mu\hbar}$, 
$\ell=0,\pm 1,\pm2,\ldots$. 
It is natural to call this phenomenon the ``collapse of quantum averages.''

The paper is organized as follows. 
We start with reviewing the general equation describing the evolution of 
quantum averages in Sections~1, 2. 
In Section~3 we discuss the case of an elliptic point, which has already 
appeared in the literature \cite{BIZ}, \cite{BBH}. 
In Section~4 we derive the explicit expressions for the evolution of 
quantum averages for the case of a hyperbolic point. 
We discuss the collapse phenomenon in Section~5 and quantum corrections 
in Section~6.

\addtocounter{section}{1}
\setcounter{equation}{0}
\subsection*{1.}
We consider a time independent polynomial Hamiltonian 
\begin{equation}\label{eq:1.1}
\begin{split}
&H(a_1^\dag ,\ldots,a_N^\dag , a_1,\ldots,a_N)
= \sum_{\ell,s} H_{\ell s} a_1^{\dag ^{\ell_1}} \cdots 
a_N^{\dag ^{\ell_N}} a_1^{s_1}\cdots a_N^{s_N}\cr
&\ell = (\ell_1,\ldots, \ell_N) \in \zed_+^N\ ,\qquad 
s= (s_1,\ldots,s_N) \in\zed_+^N\ .
\end{split}
\end{equation}
Here the creation and annihilation operators are defined as follows 
\begin{equation}\label{eq:1.2}
\begin{split}
&a_k^\dag   
= \frac1{\sqrt2} \left( x_k - \hbar \frac{\partial}{\partial x_k}\right)
\ ,\quad a_k = \frac1{\sqrt2} \left( x_k + \hbar \frac{\partial}
{\partial x_k}\right)\ ,\qquad 
k=1,\ldots, N\ ;\\
&[a_k^\dag,a_\ell] 
= - \delta_{k\ell}\hbar \ \text{ for }\ k,\ell = 1,\ldots,N\ .
\end{split}
\end{equation}
The condition 
\begin{equation}\label{eq:1.3} 
H_{\ell s} = H_{s\ell}^*\ ;\qquad s,\ell \in \zed_+^N
\end{equation}
ensures that the operator \eqref{eq:1.1} is symmetric. 
Let 
\begin{equation*}
\H (\alpha_1^*,\ldots,\alpha_N^*,\alpha_1,\ldots,\alpha_N) 
= \sum_{\ell,s} H_{\ell s} \alpha_1^{*\ell_1}\cdots \alpha_N^{*\ell_N} 
\alpha_1^{s_1} \cdots \alpha_N^{s_N}\ ,\qquad \alpha\in \cee^N\ ;
\end{equation*}
then $H$ is the Wick quantization of $\H$.
Let the total degree of $\H$ be $d$.

We remind the definition of the Poisson vectors and the set of coherent 
states \cite{G} (see also \cite{BSh}) that we need below. 

Let for $\alpha\in\cee^N$ the Poisson vector $\Phi_\alpha$ be defined 
as follows: 
\begin{equation}\label{eq:1.4} 
\Phi_\alpha (x) = (\pi \hbar)^{-N/4} \exp 
\left\{ -\frac1{2h} (x^2 - 2\sqrt{2}\, x\cdot \alpha +\alpha^2)\right\} \ ,
\qquad x\in\real^n\ .
\end{equation}
The coherent state is the normalized Poisson vector: 
\begin{equation}\label{eq:1.5}
|\alpha \rangle 
= \exp \left( - \frac{|\alpha|^2}{2\hbar}\right) |\Phi_\alpha \rangle\ .
\end{equation}
For an operator-valued function $F(t)$, the Heisenberg equation describes 
the evolution of an observable
\begin{equation}\label{eq:1.6}
\dot F = \frac{i}{\hbar} [H,F]\ .
\end{equation}
The corresponding averages are defined as follows:
\begin{equation}\label{eq:1.7}
f(\alpha^*,\alpha,t) = \langle \alpha |F(t)|\alpha \rangle\ ,\qquad 
\alpha\in\cee^N\ .
\end{equation}

\addtocounter{section}{1}
\setcounter{equation}{0}
\subsection*{2.}        
In our recent paper \cite{VB} we derived a general equation for the 
averages, extending the earlier results \cite{ST}, \cite{BBH}. 
Here we recall the form of the equation for $f(\alpha^*,\alpha,t)$
(see eqn. (1.10) in \cite{VB}):
\begin{equation}\label{eq:2.1}
\begin{split}
&\frac{\partial}{\partial t} f(\alpha^*,\alpha,t) =\\
&\qquad = \frac{i}{\hbar} \sum_{r\in\zed_+^N} \frac1{r!} 
\left\{ \left(\frac{\partial}{\partial\alpha}\right)^r \H (\alpha^*,\alpha) 
\left( \hbar \frac{\partial}{\partial\alpha^*}\right)^r 
- \left(\frac{\partial}{\partial \alpha^*}\right)^r \H (\alpha^*,\alpha) 
\left(\hbar\frac{\partial}{\partial\alpha}\right)^r\right\} 
f(\alpha^*,\alpha,t)\ .
\end{split}
\end{equation}
Here $r! = f_1!\ldots r_N!$; $(\frac{\partial}{\partial\alpha_1})^r 
= (\frac{\partial}{\partial\alpha_1})^{r_1}\cdots 
(\frac{\partial}{\partial\alpha_N})^{r_N}$; 
$(\frac{\partial}{\partial\alpha^*})^r = 
(\frac{\partial}{\partial\alpha_1^*})^{r_1} \cdots 
(\frac{\partial}{\partial\alpha_N^*})^{r_N}$. 

Clearly, only the terms with $1\le r_1 +\cdots + r_N\le d$ contribute 
to the sum in \eqref{eq:2.1}. 
We make one general remark about the initial value problem for \eqref{eq:2.1} 
such that 
\begin{equation}\label{eq:2.2} 
f(\alpha^*,\alpha,0) = \alpha^{*m} \alpha^q \ ;\qquad m,q\in\zed_+^N\ .
\end{equation}
It often happens that the coefficients in the Hamiltonian contain certain 
parameters of nonlinearity, for example, 
\begin{equation}\label{eq:2.4}
H_{\ell,s} = \mu_{|\ell +s|} G_{\ell,s}
\end{equation}
where $\mu_j$ are real, $j= 2,\ldots,d$; $|\ell+s| = \ell_1 + \cdots + \ell_N 
+ s_1 + \cdots + s_N$. 
It is of interest to ask how the solution $f(\alpha^*,\alpha,t)$ depends 
on these parameters (and on $\hbar$). 
{From} the general form of the equation \eqref{eq:2.1} it is reasonable 
to expect that the solution to \eqref{eq:2.1}, \eqref{eq:2.2} is of the form 
\begin{equation}\label{eq:2.5} 
f(\alpha^*,\alpha,t) = \hbar^{\frac{m+q}2} \F 
\left( \frac{\alpha^*}{\hbar^{1/2}} , \frac{\alpha}{\hbar^{1/2}} ,\mu_2, 
\hbar^{1/2} \mu_3, \hbar\mu_4,\ldots, \hbar^{\frac{j}2 -1} \mu_j,\ldots, 
\hbar^{\frac{d}2 -1} \mu_d, t\right)\ .
\end{equation}
The explicit form of $\F$ for a general Hamiltonian is not easy to determine. 
In what follows we discuss two examples of nonlinear systems where such a 
computation can be made. 

In the sequel we will compare solutions to \eqref{eq:2.1}--\eqref{eq:2.2} 
with the solution to Liouville's' equation of classical mechanics 
\begin{gather}
\frac{\partial}{\partial t} f_{\cl} (\alpha^*,\alpha,t) 
= i \sum_{j=1}^N \left( \frac{\partial}{\partial \alpha_j} 
\H (\alpha^*,\alpha) \frac{\partial}{\partial\alpha_j^*} 
- \frac{\partial}{\partial\alpha_j^*} \H (\alpha^*,\alpha) 
\frac{\partial}{\partial\alpha_j}\right) f_{\cl} (\alpha^*,\alpha,t)
\label{eq:2.6} \\
\noalign{\vskip6pt}
f_{\cl} (\alpha^*,\alpha,0) = \alpha^{*m} \alpha^q\ ,\qquad 
m,q\in\zed_+^N\ .\label{eq:2.7}
\end{gather}
Sometimes more general initial datum ought to be considered. 

\addtocounter{section}{1}
\setcounter{equation}{0}
\subsection*{3. Elliptic point (See \cite{BBH}, \cite{BIZ}.)}
Since this case has been already discussed in literature, we can be brief. 
Let $N=1$, 
\begin{equation}\label{eq:3.1} 
\H(\alpha^*,\alpha) = \omega |\alpha|^2 + \mu |\alpha|^4\ ,
\end{equation}
i.e., $H(a^\dag,a) = \omega a^\dag a + \mu a^{\dag2} a^2$.
The equation  \eqref{eq:2.1} takes the form 
\begin{equation}\label{eq:3.2} 
\frac{\partial}{\partial t} f(\alpha^*,\alpha,t) 
= i(\omega + 2\mu |\alpha|^2) 
\left( \alpha^* \frac{\partial}{\partial\alpha^*} 
- \alpha\frac{\partial}{\partial\alpha}\right) f (\alpha^*,\alpha,t)
+i\mu \hbar\left( \alpha^{*2} \left(\frac{\partial}{\partial\alpha^*}\right)^2
- \alpha^2 \left(\frac{\partial}{\partial\alpha}\right)^2 \right) f
(\alpha^*,\alpha,t)\ .
\end{equation}
The solution to the equation~\eqref{eq:3.2} with the initial condition 
\begin{equation}\label{eq:3.3} 
f(\alpha^*,\alpha,0) = \alpha^{*m} \alpha^q
\end{equation}
has the following form 
\begin{equation}\label{eq:3.4}
f(\alpha^*,\alpha,t) = \alpha^{*m}\alpha^q 
e^{i\omega t(m-q) + i\mu \hbar t(m(m-1)-q(q-1))} 
\  
e^{(e^{2i\mu\hbar(m-q)t}-1)\frac{|\alpha|^2}{\hbar}}\ .
\end{equation}
The  solution to Liouville's equation \eqref{eq:2.6}, \eqref{eq:2.7}
for the same Hamiltonian \eqref{eq:3.1} is 
\begin{equation}\label{eq:3.5}
f_{\cl} (\alpha^*,\alpha,t) = \alpha^{*m} \alpha^q 
e^{i(\omega +2\mu |\alpha|^2)(m-q)t}\ .
\end{equation}
Assuming $|\mu\hbar t| \ll 1$; $m\ne q$ both of order 1, we get 
\begin{equation*}
\begin{split}
f(\alpha^*,\alpha,t)  
&=f_d(\alpha^*,\alpha,t) (1+ i\mu \hbar t(m(m-1)- q(q-1) + O(|\mu\hbar t|))\\
&\qquad \cdot \exp 
(-2\mu^2 \hbar t^2 |\alpha|^2 (m-q)^2 + O (|\mu^3 \hbar^2 t^3|\, 
|\alpha|^2))\ .
\end{split}
\end{equation*}
In particular by the time $\frac1{|\mu|\,|\alpha|\sqrt{\hbar}}$ quantum 
corrections are of the same order of magnitude as the classical solution.

\addtocounter{section}{1}
\setcounter{equation}{0}
\subsection*{4. Hyperbolic point}
Let $N=1$, 
\begin{equation}\label{eq:4.1}
H(a^\dag ,a) = i\omega (a^{+2}- a^2) + \mu (a^{+2} -a)^2\ .
\end{equation}
In this section we give an explicit solution to \eqref{eq:2.1}. 
The operator \eqref{eq:4.1} corresponds to the Wick Hamiltonian 
\begin{equation}\label{eq:4.2} 
\H (\alpha^*,\alpha) = i\omega (\alpha^{*2} - \alpha^2) 
+ \mu (\alpha^{*2} - \alpha^2)^2 
- 4\mu \hbar \alpha^* \alpha - 2\mu \hbar^2\ .
\end{equation}
this Hamiltonian contains hyperbolic point at the origin with 
Lyapunov exponents 
\begin{equation}\label{eq:4.3} 
\lambda_\pm = \pm 2\sqrt{\omega^2 - \mu^2 \hbar^2} \ ,\qquad 
|\mu|\hbar < |\omega|\ .
\end{equation}
For simplicity we consider the initial condition for \eqref{eq:2.1} of the 
form 
\begin{equation}\label{eq:4.4} 
f(\alpha^* ,\alpha,0) = \langle \alpha |\hat x^n|\alpha\rangle\ ,
\qquad n=1,2,3,\ldots\ .
\end{equation}
This corresponds to quantum averages in coherent states of the $n$-th 
power of the coordinate operator. 

An arbitrary polynomial in $\alpha^*,\alpha$ can be handled by the same 
method, but the resulting expression is of more involved combinatorial 
structure. 

It will be convenient to introduce the coordinate and the 
momentum  operator 
\begin{equation}\label{eq:4.5} 
\hat x = \frac1{\sqrt2} (a^\dag +a)\ ,\quad 
\hat p = \frac{i}{\sqrt2} (a^\dag -a)
\end{equation} 
and their evolution according to Heisenberg's equation \eqref{eq:1.6}
\begin{align}
X(t) & = e^{\frac{i}{\hbar} Ht} \hat x e^{-\frac{i}{\hbar} Ht}\ ,
\label{eq:4.6}\\
P(t) & = e^{\frac{i}{\hbar} Ht} \hat p e^{-\frac{i}{\hbar} Ht}\ .
\label{eq:4.7}
\end{align}
Then $X(0) =\hat x$, $P(0) =\hat p$, $[X(t),P(t)] = i\hbar$. 
The operator $H$ can be expressed using \eqref{eq:4.1}, \eqref{eq:4.5} as 
\begin{equation}\label{eq:4.8}
H = 2\omega \hat x\hat p - i\omega \hbar  + \mu (2i\hat x\hat p +\hbar)^2\ .
\end{equation}
Since $[H,\hat x\hat p] =0$, it follows that 
\begin{equation}\label{eq:4.9} 
X(t)\  P(t) = \hat x\hat p\ ,\qquad t\in\real\ .
\end{equation}
This obvious remark will be crucial for our computation. 
We have from \eqref{eq:1.6}, \eqref{eq:4.8}, \eqref{eq:4.9} 
\begin{align}
\dot X (t) & = (2\omega - 8\mu  X(t)P(t))X(t) 
= (2\omega - 8\mu \hat x\hat p) X(t)
\label{eq:4.10}\\
\dot P (t) & = (-2\omega - 8\mu i\hbar + 8\mu X(t) P(t)) P(t) 
= (-2\omega - 8\mu i\hbar + 8\mu \hat x\hat p) P(t) \ .
\label{eq:4.11}
\end{align}
Using \eqref{eq:4.10}, \eqref{eq:4.11} and induction in  $n$, we get 
\begin{align}
\frac{d}{dt} X^n (t) 
& = (2n\omega - 4\mu i\hbar n(n-1) - 8n\mu \hat x\hat p) X^n (t)
\label{eq:4.12}\\
\frac{d}{dt} P^n (t) 
& =  (-2n\omega - 4\mu i\hbar n(n+1) + 8n\mu \hat x\hat p) P^n (t) \ .
\label{eq:4.13}
\end{align}
{From} \eqref{eq:4.12}, formally
\begin{equation}\label{eq:4.14}
\begin{split} 
X^n (t) 
& = e^{(2n\omega - 4\mu i\hbar n(n-1) - 8n\mu \hat x\hat p)t} \hat x^n\\
& = e^{(2n\omega -4\mu i\hbar n(n-1)) t} 
e^{8in\mu\hbar t\hat x\frac{\partial}{\partial x}} \hat x^n\ .
\end{split}
\end{equation}

For any entire function $G(x)$, $x\in \cee$ we have 
\begin{equation}\label{eq:4.15}
\left(e^{i\tau x\frac{\partial}{\partial x}} G\right) (x) 
= G(e^{i\tau} x)\ .
\end{equation}
But the coherent state $|\alpha \rangle$ is analytic in $x$, therefore
using \eqref{eq:4.14}, \eqref{eq:4.15} 
\begin{equation}\label{eq:4.16}
\begin{split}
&  f(\alpha^*,\alpha,t) = \langle \alpha |X^n(t)|\alpha\rangle\\
&\qquad  = 
e^{-\frac{|\alpha|^2}{\hbar} + 2\omega n t + 4\mu i\hbar n(n+1)t} 
\int \Phi_\alpha^* (x)  x^n \Phi_\alpha (e^{8in\mu\hbar t} x)\,dx \\
&\qquad = 
\frac1{(\pi\hbar)^{1/2}} e^{-\frac{(\alpha +\alpha^*)^2}{\hbar}  }
e^{2\omega n t + 4\mu i\hbar tn(n+1)} 
\int_{-\infty}^\infty x^n 
e^{-\frac1{2\hbar} [(1+e^{16\mu in\hbar t}) x^2 
- 2\sqrt2 (\alpha^* + \alpha e^{8\mu in\hbar t}) x]}
\, dx \ .
\end{split}
\end{equation}
We compute the integral in the right side of \eqref{eq:4.16} assuming
\begin{equation}\label{eq:4.17}
t\ne \frac{\pi}{16\mu n\hbar} + \ell \frac{\pi}{8\mu\hbar n}\ ,\qquad 
\ell\in\zed\ .
\end{equation}
The condition \eqref{eq:4.17} will be discussed in detail in the next 
section; here we just note that \eqref{eq:4.17} implies \break
$\Re (1+e^{16\mu in\hbar t}) >0$ and the integral in \eqref{eq:4.16} is 
absolutely convergent. 
Changing variables in \eqref{eq:4.16} we arrive at the expression 
\begin{equation}\label{eq:4.18}
\begin{split}
&f(\alpha^*,\alpha,t) = \\
&= \frac{(2\hbar)^{\frac{n+1}2} e^- \frac{(\alpha+\alpha^*)^2}{2\hbar}}
{(\pi\hbar)^{1/2}} 
e^{2\omega tn}  
\left( 
\frac{e^{4\mu i\hbar tn}}
	{(1+e^{16\mu in\hbar t})^{1/2}}
\right)^{n+1}\\
&\qquad 
e^{\frac{(\alpha^* e^{-4in\mu\hbar t} + \alpha e^{4in\mu\hbar t})^2}
{2\hbar \cos 8n\mu\hbar t}} 
\int_{-\infty}^\infty 
\left( x + \frac{\alpha^* + \alpha e^{8\mu in\hbar t}}
{\sqrt{\hbar} (1+e^{16\mu in\hbar t})^{1/2}} \right)^n 
e^{-x^2}\, dx \ .
\end{split}
\end{equation}
In \eqref{eq:4.18} we choose the branch of the square root so that 
$\Re \sqrt{1+ e^{16\mu in\hbar t}} >0$. 
The expression $e^{4\mu i\hbar tn} / \sqrt{1+e^{16\mu in\hbar t}}$ 
because of this condition ought to be interpreted as follows: 
\begin{equation}\label{eq:4.19} 
\frac{e^{4\mu i\hbar tn}}
	{\sqrt{1+e^{16\mu i\hbar tn}} }
= \frac1{\sqrt{2\cos 8\mu\hbar tn}} 
\defeq 
\frac1{\sqrt{2|\cos 8\mu \hbar tn|}} i^{\left[\frac12 
+ \frac{8\mu n\hbar}{\pi} t\right]}\ .
\end{equation}
The square bracket in \eqref{eq:4.19} stands for the entire part of a 
real number. 
In other words, passing through every point \eqref{eq:4.17} from the left 
to the right contributes a factor of $e^{\pi i/4}$ to this expression. 

Therefore,
\begin{equation}\label{eq:4.20}
\begin{split}
&f(\alpha^*,\alpha,t) =\\
&\qquad  = \frac{\hbar^{n/2}}{\sqrt{\pi}} 
e^{-\frac{(\alpha +\alpha^*)^2}{2\hbar}} 
\frac{e^{2\omega nt}} {(\cos 8\mu n\hbar t)^{\frac{n+1}2}}
e^{\frac{(\alpha^* e^{-4in\mu\hbar t} + \alpha e^{4in\mu \hbar t})^2}
	{2\hbar \cos 8n\mu\hbar t}}\\
&\qquad\qquad  \sum_{k=0}^{[n/2]} 
\frac{n!}{(2k)!(n-2k)!} 
\frac{(\alpha^* e^{-4in\mu\hbar t} + \alpha  e^{4in\mu\hbar t})^{n-2k}}
	{\hbar^{\frac{n}2 -k} (2\cos 8n\mu\hbar t)^{\frac{n}2 -k}} 
\int_{-\infty}^\infty x^{2k} e^{-x^2} \,dx \\ 
&\qquad = e^{-\frac{(\alpha +\alpha^*)^2}{2\hbar}} 
\frac{e^{2\omega nt}} {(\cos 8\mu n\hbar t)^{\frac{n+1}2}} 
e^{\frac{(\alpha^* e^{-4in\mu\hbar t}+ \alpha e^{4in\mu\hbar t})^2}
	{2\hbar \cos 8n\mu\hbar t}}\\
&\qquad\qquad  
2^{-n/2} \sum_{k=0}^{[n/2]} 
\frac{n! (2k-1)!!}{(2k)! (n-2k)! 2^k}  2^k \hbar^k 
\frac{(\alpha^* e^{-4in\mu\hbar t} + \alpha e^{4in\mu\hbar t})^{n-2k}}
	{(\cos 8n\mu\hbar t)^{\frac{n}2 -k}} \\
&\qquad = e^{-\frac{(\alpha +\alpha^*)^2}{2\hbar}} 
\frac{e^{2\omega nt}} {(\cos 8\mu n\hbar t)^{\frac{n+1}2}}
e^{\frac{(\alpha^* e^{-4in\mu\hbar t} + \alpha e^{4in\mu\hbar t})^2}
	{2\hbar \cos 8n\mu\hbar t}} \\
&\qquad\qquad  
2^{-n/2} \sum_{k=0}^{[n/2]} 
\frac{n!}{2^k k! (n-2n)!} \hbar^k 
\frac{(\alpha^* e^{-4in\mu\hbar t} + \alpha e^{4in\mu \hbar t})^{n-2k}}
	{(\cos 8\mu n\hbar t)^{\frac{n}2-k}}\ .
\end{split}
\end{equation}
The identity \eqref{eq:4.20} is the central result of this paper.

The explicit form of the equation \eqref{eq:2.1} for the 
Hamiltonian \eqref{eq:4.2} is as follows: 
\begin{equation}\label{eq:4.21}
\begin{split}
\frac{\partial}{\partial t} f &=
i\left[ -2i\omega \alpha - 4\mu (\alpha^{*2} -\alpha^2)\alpha 
- 4\mu\hbar \alpha^*\right] \frac{\partial}{\partial \alpha^*} f\\
&\qquad
- i \left[ 2i\omega\alpha^* + 4\mu (\alpha^{*2}-\alpha^2)\alpha^* 
- 4\mu\hbar \alpha\right] \frac{\partial}{\partial \alpha} f\\
&\qquad
+ i\hbar \left[ - i\omega -2\mu (\alpha^{*2}-3\alpha^2)\right] 
\left(\frac{\partial}{\partial\alpha^*}\right)^2 f\\
&\qquad
- i\hbar \left[ i\omega +2\mu (3\alpha^{*2} - \alpha^2)\right] 
\left( \frac{\partial}{\partial \alpha}\right)^2 f\\
&\qquad
+ 4i\hbar^2 \mu\alpha \left(\frac{\partial}{\partial \alpha^*}\right)^3 f
- 4i\hbar^2 \mu\alpha^* \left(\frac{\partial}{\partial\alpha}\right)^3 f\\
&\qquad
+ i\hbar^3 \mu \left(\frac{\partial}{\partial\alpha^*}\right)^4 f
- i\hbar^3 \mu \left(\frac{\partial}{\partial\alpha}\right)^4 f
\end{split}
\end{equation}

Verifying that the expression \eqref{eq:4.20} satisfies \eqref{eq:4.21} 
directly, e.g., for $n=1$ is a very long computation, but it does indeed. 

\addtocounter{section}{1}
\setcounter{equation}{0}
\subsection*{5. Collapse of quantum averages}
To simplify matters we take $n=2$. 
We have using \eqref{eq:4.20} 
\begin{equation*}
f(\alpha^*,\alpha,t) 
= \frac12 e^{-\frac{(\alpha+\alpha^*)^2}{2\hbar}} 
\frac{e^{4\omega t}} {(\cos 16\mu\hbar t)^{3/2}}
e^{\frac{(\alpha^* e^{-8i\mu \hbar t} +\alpha e^{8i\mu\hbar t})^2}
{2\hbar\cos 16\mu\hbar t}} 
\left[\frac{(\alpha^* e^{-8i\mu\hbar t} + \alpha e^{8i\mu\hbar t})^2}
	{\cos 16\mu\hbar t} + \hbar \right]\ .
\end{equation*}
We discuss the limit of this expression as $t\to \frac{\pi}{32\mu\hbar} -0$. 
For simplicity let $\alpha =i$. 
Since 
\begin{equation*}
(\alpha^* e^{-8i\mu\hbar t} + \alpha e^{8i\mu\hbar t})^2 
= \left( -i \left( \frac{1-i}{\sqrt2}\right) + 
i \left(\frac{1+i}{\sqrt2}\right) \right)^2 
=2\ \text{ at }\ t = \frac{\pi}{32\mu\hbar}\ ,
\end{equation*}
we have 
\begin{equation*}
f(-i,i,t)\to \infty\ \text{ as }\ t\to \frac{\pi}{32\mu\hbar} -0\ .
\end{equation*}
But 
\begin{equation*}
f(\alpha^*,\alpha,t) = \|\hat x e^{-\frac{i}{\hbar} Ht} |\alpha\rangle
\|_{L^2(\real)}^2\ .
\end{equation*}
The phenomenon we observed is the quantum evolution of coherent 
states  for certain $\alpha$
in our explicitly integrable system may take them out of the 
{\em domain\/} of the self-adjoint {\em unbounded\/} operator $\hat x$ 
in $L^2(\real)$. 
This happens along a discrete sequence 
\begin{equation*}
t_\ell = \frac{\pi}{32\mu\hbar} + \ell \frac{\pi}{16\mu \hbar}\ ,\qquad 
\ell = 0,\pm1,\pm2,\ldots\ .
\end{equation*}
It is natural  to describe it as collapse of certain quantum averages 
that occurs at these times $t_\ell$. 

\addtocounter{section}{1}
\setcounter{equation}{0}
\subsection*{6. Quantum corrections in presence of a hyperbolic point}
For simplicity we set $n=1$ in \eqref{eq:4.20}, i.e., 
\begin{equation}\label{eq:6.1} 
f(\alpha^*,\alpha,t) = 
\langle \alpha |X(t)|\alpha\rangle  
=\frac1{\sqrt2} 
e^{-\frac{(\alpha+\alpha^*)^2}{2\hbar}} 
\frac{e^{2\omega t}} {(\cos 8\mu\hbar t)^{3/2}}
e^{\frac{(\alpha^* e^{-4i\mu\hbar t} + \alpha e^{4i\mu\hbar t})^2}
{2\hbar \cos 8\mu \hbar t}} 
(\alpha^* e^{-4i\mu \hbar t} + \alpha e^{4i \mu\hbar t}) \ .
\end{equation}
We assume as in Section 3 that $|\mu\hbar t| \ll1$. 
A simple computation leads to the result 
\begin{equation}\label{eq:6.2}
\begin{split}
f(\alpha^*,\alpha,t) 
& = \frac1{\sqrt2} e^{2\omega t+ 4i\mu t(\alpha^2-\alpha^{*2})} 
\left(1+ \O (\mu^2 \hbar^2 t^2)\right)\\
\noalign{\vskip6pt}
&\qquad \times \left( (\alpha^* + \alpha) + 4i\mu\hbar t(\alpha-\alpha^*) 
+ \O (|\alpha| \mu^2 \hbar^2 t^2) \right)\\
\noalign{\vskip6pt}
&\qquad \times  \exp \left( 16 \mu^2 \hbar t^2 |\alpha|^2 
+ \O (|\alpha|^2 |\mu^3 \hbar^2 t^3|) \right) \ .
\end{split}
\end{equation}
It is of interest to note that 
\begin{equation*}
f_{\cl} (\alpha^* ,\alpha,t) = \frac1{\sqrt2} 
e^{2\omega t+4i\mu t(\alpha^2 -\alpha^{*2})} 
(\alpha^* +\alpha)\ .
\end{equation*}
Over the time of order $\frac1{|\mu|\,|\alpha|\sqrt{\hbar}}$ the 
quantum corrections are of the same order of magnitude as classical solutions.
To measure the deviation of the quantum average from its classical value 
the most natural quantity is dispersion. 
We will give estimates for 
\begin{equation*}
D(\alpha^*,\alpha,t) = \langle \alpha |X^2 (t)|\alpha \rangle 
- \langle \alpha |X(t)|\alpha\rangle^2
\end{equation*}
which is the dispersion of $X(t)$ in coherent state. 
It is a purely quantum quantity; the corresponding classical quantity 
made out of solution to the Liouville's equation of classical mechanics 
vanishes identically. 
To present the result in a simpler way we assume 
\begin{equation}\label{eq:6.3} 
|\mu \hbar t| \ll 1\ ,\quad 
|\alpha|^2 \gg \hbar\ ,\quad 
\mu^2 \hbar t^2 |\alpha| \ll 1\ .
\end{equation}
The main term in the dispersion $D(\alpha^*,\alpha,t)$ under these 
assumptions is as follows: 
\begin{equation}\label{eq:6.4}
D(\alpha^*,\alpha,t) \approx 
e^{4\omega t - 8i\mu\hbar t(\alpha^2-\alpha^{*2})} 
\left( \frac12 \hbar + 4i\mu\hbar t(\alpha^2 - \alpha^{*2}) 
+ 16 (\alpha^* + \alpha)^2 \mu^2 \hbar t^2 |\alpha|^2\right)\ .
\end{equation}
If we replace \eqref{eq:6.3} by 
\begin{equation}\label{eq:6.5}
|\mu \hbar t| \ll 1\ ,\quad 
|\alpha |^2\gg \hbar\ ,\quad 
\mu^2 \hbar t^2 |\alpha|^2 \gg 1\ ,
\end{equation}
we get the expression 
\begin{equation}\label{eq:6.6}
D(\alpha^*,\alpha,t) \approx \frac12 e^{4\omega t+8i\mu (\alpha^2-\alpha^{*2})}
(\alpha^* +\alpha)^2 
\exp (64 \mu^2 \hbar t^2 |\alpha|^2)\ .
\end{equation}
Finally, if 
\begin{equation}\label{eq:6.7} 
|\mu\hbar t|\ll 1\ ,\quad 
|\alpha|^2 \gg \hbar\ ,\quad 
64 \mu^2 \hbar t^2 |\alpha|^2 \approx 1
\end{equation}
we get 
\begin{equation}\label{eq:6.8} 
D(\alpha^*,\alpha,t) \approx \frac12 
e^{4\omega t+8i\mu (\alpha^2 -\alpha^{*2})} 
(\alpha^* +\alpha)^2 
\left[ \exp (64 \mu^2 \hbar t^2|\alpha|^2) 
- \exp (32\mu^2 \hbar t^2 |\alpha|^2\right)]\ .
\end{equation}
It is clear from these expressions that over a logarithmically small time 
$C\log \frac1{\hbar}$ the quantum dispersion becomes at least of order~1.

\subsection*{Conclusions}

For the polynomial Hamiltonian of degree 4 in $a^\dag,a$ with a hyperbolic 
point at the origin we here obtained the explicit formulae for quantum 
averages in coherent states valid for all times. 
Quantum corrections for averages become significant already on a 
logarithmically small time (in $\hbar$). 
The quantum averages for simple observables, such as $\hat x^n$, 
blow up along a discrete sequence of times for this Hamiltonian. 

\subsection*{Acknowledgements}

This work was supported by the Department of Energy under the contract
W-7405-ENG-36 and DOE Office of Basic Energy Sciences.  
The work of G.P.B. was partly supported by the National Security Agency 
(NSA) and by the Advanced Research and Development Activity (ARDA). 
The work of M.V. was supported in part by the National Science Foundation 
Grant DMS-0301531.
The authors thank Margaret Combs for her excellent typing of the manuscript.

\bibliography{longtime}

\end{document}